# Microwave impedance readout of a hafnium microbridge detector

A.V. Merenkov[1], V.I. Chichkov[1], A.B. Ermakov[1,2], A.V. Ustinov[1,3], S.V. Shitov[1,2]

[1] *National University of Science and Technology MISIS, 119049, Moscow, Russia*
[2] *Kotel'nikov Institute of Radio Engineering and Electronics, 125009, Moscow, Russia*
[3] *Physikalisches Institut, Karlsruhe Institute of Technology, 76131 Karlsruhe, Germany*

We present proof-of-operation for a new method of electron thermometry using microwave impedance of a hafnium micro-absorber. The new method leads to an ultimate THz-range detector suitable for microwave readout and frequency division multiplexing. The sensing part of the device is a hot-electron-gas absorber responding to the incident radiation by variation of its impedance measured at probing frequency about 1.5 GHz. The absorber is a microbridge made from hafnium ($T_c \approx 375$ mK, $R_N \approx 30$ Ω) sized 2.5 μm by 2.5 μm by 50 nm and integrated with a planar 600-700 GHz antenna placed near the *open* end of a quarter-wave CPW resonator ($Q$-factor $\sim 10^4$). All elements of the circuit, except the microbridge, are made from 100-nm thick Nb, including the resonator, which is weakly coupled to a throughput line. The device was tested at 50-350 mK smoothly responding with its transmission coefficient $|S_{21}|$ to applied microwave power at the resonance frequency. We have found that the power absorbed by the bridge fits to the model of hot electron gas, $P \sim T_e^n - T_{ph}^n$ ($n = 5\ldots6$). The idle *NEP* down to $\approx 10^{-18}$ W/√Hz and the corresponding cross-over temperature for photon background $\approx 5$ K are estimated from the measured data. The saturation power of about 1 pW and possibility of moderate gain are anticipated for a practicable device operating at temperature 200 mK. Since the optimum readout frequency is found exactly at the resonance, the detector is insensitive to most phase instabilities at the probing frequency.

The superconducting THz-range imaging detectors are rapidly developing field. One of the matured detector technologies is the superconducting Transition Edge Sensor (TES), which is a bolometer operating at ultra-low temperatures. Incident radiation heats a suspended membrane absorber[1-3] while TES thermometer, attached to the absorber, indicates the growing temperature. The Frequency Division Multiplexing (FDM)[4-5] technology is widely used for large TES arrays. However, few specific drawbacks of TES bolometers, such as limited mechanical stability of the membrane, spurious response to the cosmic rays, poor array packaging, and slow response are known. Studies on non-equilibrium electrons in solids motivated invention of a few membrane-free detectors. One of already proven technologies is the Microwave Kinetic Inductance Detector (MKID), which demonstrated Noise Equivalent Power (*NEP*) down to $\approx 10^{-19}$ W/√Hz[6-8]. The MKID can respond to excitation of few quasiparticles (QP's) with resonance frequency and Q-factor of a superconducting resonator[9] operating at temperatures much below its critical temperature, $T_c$. The MKID responds by a pair-breaking process triggered by high-energy photons. The choice of the MKID technology for array applications is often motivated by convenience of its *microwave* FDM circuit.

The effect of electron gas (e-gas) absorption[10-14] is another solution for the terahertz-range membrane-free *bolometers*. Due to weak electron-phonon interaction at low temperatures, the electron subsystem of many conducting materials[10] can be heated by electrical current from an antenna and reaches its internal thermal equilibrium at a temperature $T_e$, which is higher than the lattice temperature $T_{ph}$. The concept of a superconducting E-Gas Absorber (EGA) assumes that the number of QP's, unlike the MKID case, must be large. For this reason the ultimate *NEP* of EGA is governed by thermal fluctuations in the e-gas and by the number of absorbing QP's. So the thermal isolation of the EGA can be improved by reducing the volume of the absorber. However, the measurement of the electron temperature is the most challenging task.

For superconductors, the e-gas effect becomes pronounced near $T_c$ at temperatures below 1 K, for example, below $T_c \approx 0.4$ K for Ti and Hf[12]. Weak signals can be detected with EGA by measuring variations of its thermal noise or resistance[11-14]. It is also possible to measure the current of hot (non-equilibrium) electrons escaping from a normal-metal absorber through the tunnel barrier of SIN junctions[15-16]. It was demonstrated that THz current from an integrated antenna can rise the electron temperature and resistance of a small-volume superconducting bridge[13-14]. This concept promises *NEP* down to and even below $\approx 10^{-20}$ W/√Hz[13] that is so far the lowest reported level for any direct detector. It is worth to note here that probing of EGA using dc or low frequency is rather difficult technology, which requires weakly perturbing measurements of low resistance using multi-stage EMF-filters, not to mention need in costly SQUID-amplifiers. This is why *microwave control and readout* of an EGA-based bolometer seems a promising way towards an ultimate detector. In this Letter, we report on proof-of-operation for a new method of reading the electron temperature of a superconducting THz-range detector exploiting the effect of the non-linear electron-gas absorption at microwave frequencies.

On the one hand, it is feasible treating EGA as a thermistor[17-20] $\sim 1$ Ω most sensitive near $T_c$. On the other hand, a general problem exists in the case of the superconducting "thermistor" at microwave frequencies. The gap energy, Δ, of a superconductor $\Delta(T) \to 0$ near $T_c$ that means a very high probability of breaking Cooper pairs by the probing photons. If the probing frequency, $f$, yeilds $hf$

> $\Delta(T_c)$, the bridge behaves as a normal conductor exhibiting nearly constant impedance $\approx R_N$. The feasibility of a microwave-readable superconducting "thermistor" was preliminary analyzed in Ref. [22].

*The temperature-dependent impedance* of a superconductor, $Z(T_{bath})$, can be described with the Mattis-Bardeen theory[23] (MBT), which assumes the following *equilibrium* condition, $T_{bath} = T_{ph} = T_e$. The number of QP's and associated microwave loss are defined by *equilibrium* temperature, $T_{ph}$. We assumed that the electrical loss depends on the electron temperature, and put forward the *hypothesis of equality*: the impedance remains the same in the equilibrium, $T_{bath1} = T_{ph1} = T_{e1}$, and in the e-gas regime, $T_{bath2} \approx T_{ph2} << T_{e2}$, if $T_{e1} = T_{e2}$. This approach can be justified due to slow electron-phonon interaction, when few long-living QP's can be quickly excited by microwaves up to the energy corresponding to temperature $T_{ex} >> T_{ph}$. A quick energy exchange within e-gas eventually leads to temperature *equilibrium* within whole electron sub-system acquiring higher effective temperature $T_e > T_{ph}$. Using the above hypothesis, $Z(T_e)$ can be predicted using MBT[22].

*Comparing the predicted and experimental* $Z(T_e)$ is possible by defining parameters $S_{21}$ and $S_{31}$ of a device presented schematically in the right inset of Fig. 1. The $S_{31}$ parameter governs the EGA-coupled probing power, $P$. The exponent $n$ in the relation $P = k_n(T_e^n - T_{ph}^n)$ describes the heat exchange between electrons and phonons[10-14]. The exponent $n = 4$ means dominance of the Kapitsa thermal resistance, so the unwanted phonon mediation[24-25] and larger heat loss (low thermal conductance $G$) must be anticipated. The exponent $n = 6$ corresponds to the e-gas model meaning very low electron-phonon coupling. The coefficient $k_n$ combines material parameters and geometry of the sample.

*Previously, when prototyping the readout circuit,* we avoided complexity of measurements at ultra-low temperatures. The *proof-of-concept circuits*[17-20] were tested at temperatures 1.5…5 K and readout frequencies 5…8 GHz. The prototype bridges were made from a 16-nm film of Nb providing reduced transition temperature $\approx$ 5.5 K while the rest of the circuit was made from thicker 200-nm Nb ($T_c \approx$ 9 K). The prototypes were tested for optical *NEP* using blackbodies[17] and the FDM capability was demonstrated using the seven-pixel array[18]. The FDM resonators were weakly coupled to a common throughput coplanar waveguide (CPW) line, similar to MKID[6-8]. To supply THz photons, the absorber was connected in series for both microwave and THz currents, as presented in the right inset of Fig. 1. The antenna was designed for the band 600-700 GHz that is above the Plank's peak of the Cosmic Microwave Background Radiation (CMBR). The optimum load of the antenna was 25-35 Ω that is about $R_N$ of the absorbing bridge. The optical *NEP*~$10^{-14}$ W/√Hz was demonstrated at 1.5 K confirming the EM-design. Suggesting the readout circuit, we have called it RFTES emphasizing the microwave (RF) technique of reading the impedance over the superconducting transition edge.

Here, we are presenting the most advanced RFTES version with data measured at millikelvin temperatures. To justify our present choice of RFTES/EGA combination, a brief comparison to other superconducting detectors and absorbers is given below.

*Comparison to an MKID detector.* At first glance, RFTES is similar to MKID. However, they use different microwave coupling scheme. The MKID reacts to the photon-excited QP's, which arrive from *superconducting* antenna and load the *shorted* end of the *electrically* coupled resonator with small impedance ~ $10^{-3}$ Ω. This means that a normal-metal antenna, which can be better at THz, should insert too large loss if integrated with the resonator. The RFTES uses a *magnetically* coupled resonator, and the bridge is loading the resonator near its *open end*[17-19] thus minimizing the possible effect from a normal-metal antenna. To avoid the dark QP-activation, the probing power of an MKID has to be sufficiently *low* that unavoidably limits its saturation level. In contrast, the microwave power for RFTES must be *high enough* to activate QP's up to electron temperatures ~ $T_c$.

*Comparison to an TES bolometer.* The RFTES reacts to THz current from an integrated antenna, while a classical TES sensor receives heat defusing from a suspended absorber. Similar to TES, the Electro-Thermal Feedback[2] (ETF) is present by the shunting effect from the embedding impedance of the integrated resonator. Such integration can provide much better phase stability at higher probing frequencies and wider range of absorber impedance when compared with a dc/low-frequency wire circuit of TES. The operation temperature of the RFTES, $T_{bath}$, can be lower than $T_c$, since $T_e$ is controlled by the readout/bias power. This resembles the self-heating regime of TES, $T_{bath} < T_c$, as described, for example, in Ref. [21].

We employed *mask-free fabrication process* by patterning our devices with Heidelberg µPG-501 laser writer. Nine devices were fabricated on a 500-µm silicon wafer sized 15 mm by 15 mm. Both Hf and Nb films were deposited using *dc* magnetron sputtering. Hf bridges 2.5 µm by 6 µm were formed using the lift-off process. The 100-nm film of Nb was deposited forming terminals of the Hf bridges. The 4 mm by 4 mm chips were wire-bonded to a PCB and tested in a dilution refrigerator[26] at temperatures 50-350 mK. The microwave probing/bias power was supplied from a network analyzer using cryogenic coaxial cables. To suppress the 300-K noise, the probing power was attenuated by 40 dB at 1-K stage. The readout path included a cryogenic circulator and a low-noise amplifier (LNA) at 3-K stage of the refrigerator. Figure 1 illustrates the set of experimental data on transmittance minima, which are shown in the left inset.

*Extracting impedance of the absorber.* An equivalent scheme of the device was used as described in Ref. [22]. The active part of the impedance, $R_B = Re(Z(T_e))$, is a parameter of the scheme, which response, $S^{eq}_{21}(f,R_B)$, must fit the experimental $S_{21}(f,T_e)$. The Microwave Office software[27] was used for the simulations. Unfortunately, this Letter has no room for detailed description of the fitting procedure. In brief, $S^{eq}_{21}$ and $S^{eq}_{31}$ are related via the EM-model, and the fit defines $R_B$ and the absorbed power. We have succeeded to recover $R_B$ in the range 0.01…8 Ω; larger values are found inaccurate due to growing errors of the non-linear fitting.

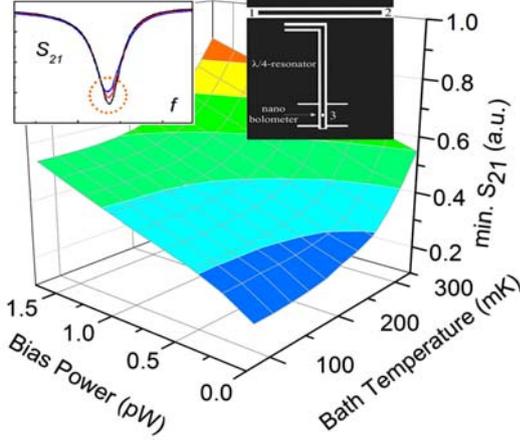

FIG 1. Three-dimensional presentation of experimental data at bias frequency 1.5 GHz. The left inset illustrates smooth control over transmission and definition for "min. $S_{21}$". The right inset shows simplified circuit layout indicating three EM-ports (metal shown in black).

It was found that the data on $R_B(T_e) = R_B(P_{bias}, T_{bath})$ can be extrapolated towards the reference value $R_N \approx 30$ Ω at $T_{bath} = T_c$ using the dependence $R_B \sim R_N \times \exp[\alpha(T-T_c)/T]$. This important result demonstrates a possibility of microwave thermometry by EGA that is a complimentary method to the TES thermometry. Following our hypothesis, the condition $T_e = const$ means $S_{21} = const$, and the curves $P_{bias}(T_{bath}, T_e = const)$ can be extracted from the data presented in Fig. 1. To define $T_e$, the equation for power balance of the EGA ($n = 6$) can be written as follows:

$$P_{bias} = \Sigma \cdot V \cdot (T_e^6 - T_{bath}^6)$$
$$G(T_e) = dP_{bias}/dT_e = 6 \cdot \Sigma \cdot V \cdot T_e^5 \quad (1)$$

Since the experimental thermal conductance, $G$, was estimated as small as $\sim 10^{-11}$ W/K, the Kapitsa thermal conductance, which is about two orders smaller, is not included in (1) and $T_{bath} \approx T_{ph}$ was assumed. The diffusion out from the bridge was neglected due to both small mean free path of QP's in the Hf film (< 2 μm) and the Andreev reflection[28] at N-S interface. The total volume of the bridge was including its overlap with the Nb electrodes yielding $V = 0.75 \cdot 10^{-18}$ m$^3$. Fitting equations (1) to experimental data is illustrated in Fig. 2 using material parameter $\Sigma = 17.5 \cdot 10^8$ W/(m$^3 \cdot$K$^6$). We have to admit that above value of $\Sigma$ is about twice larger than that in Ref.[12]. The difference can be explained by underestimation of the total loss in the circuit that, according to the equivalent scheme, leads to essential overestimation of both parameter $\Sigma$ and the loss in the bridge. Resuming, the fit of several curves using Eq. (1) for $n = 6$ is good argument in favor of both the model of microwave EGA and the abovementioned *hypothesis of equality* for the experimental bridge.

*Estimating NEP of the detector*, we assume that dominant noise in the circuit is the thermal fluctuations of the electron gas. Then the *idle, or internal*, $NEP_0$ can be calculated with the basic bolometric formula substituting the absorber temperature with the temperature of electron subsystem: $NEP_0 = \sqrt{4k_B T_e^2 G}$. The thermal conductance, $G(T_e)$, was defined independently from Eq. (1) as $G \approx 1.92 \cdot 10^{-11}$ W/K at electron temperature $T_e = 300$ mK that yields $NEP_0 \approx 1 \cdot 10^{-17}$ W/√Hz. This $NEP_0$ does not include noise of the LNA and photon background noise.

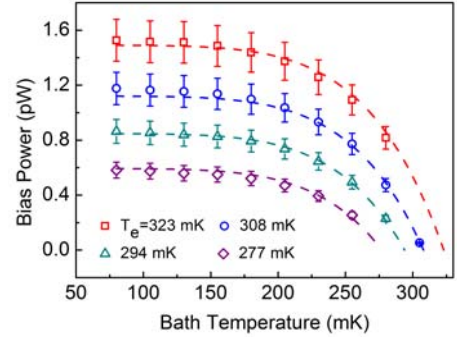

FIG 2. Fit of power exchange within electron-phonon model $P \sim T_e^6 - T_{ph}^6$ to experimental data from Fig. 1 using method of steady loss, $S_{21} = S_{21}(T_e) = const$. Values of $T_e$ are fitted for each dashed curve.

*The photon background noise* limits sensitivity to $NEP_x = \sqrt{2 P_x h f_x}$, where $P_x$ is an integral of spectral density of a black-body radiation within the band of the single-polarized antenna. According to Plank's formula

$$P_x(T) = \int_{f_1}^{f_2} \frac{2hf}{\exp(hf/kT)-1} df \quad (2).$$

Where $f_1 = 600$ GHz to $f_2 = 700$ GHz in our case. Considering possible space application of the EGA device, the CMBR at 2.8 K yields the photon noise $P_{xCMBR} \approx 6.5 \times 10^{-16}$ W and $NEP_{xCMBR} \approx 5.4 \times 10^{-19}$ W/√Hz. Since the bias power in our experiment is about three orders larger than $P_{xCMBR}$, the saturation by photons is not the case. For balloon applications, however, the detector could saturate with the sky warmer than 10 K. The Eq. (2) yields for the 1.5-GHz photons $NEP_{xbias} \approx 1.7 \times 10^{-18}$ W/√Hz. This is much lower than intrinsic $NEP_0$, however this bias-contributed noise is essentially larger than the $NEP_{xCMBR}$.

*Improvement of $NEP_0$ is possible* using smaller bridges fabricated with medium-resolution electron-beam lithography. For example, reducing volume to 0.25 μm by 0.25 μm by 25 nm ($V = 1.56 \cdot 10^{-21}$ m$^3$) should not be a problem. Such scaling is a proven technique validated for a wide range of geometries, and model (1) should work similarly to Ref.[12]. Assuming the material parameters unchanged, the idle $NEP_0 \approx 6 \times 10^{-19}$ W/√Hz can be estimated that is about equal to $NEP_{xCMBR}$. The noise of microwave bias photons gives $NEP_{xbias} \approx 1.2 \times 10^{-19}$ W/√Hz at $P_{bias} \approx 7 \times 10^{-15}$ W. This implies that the smaller bridge will remain not saturated by CMBR.

*Estimating NEP contribution from LNA*, the power-to-power conversion is studied using experimental data. The equivalents of signal power, $P_{in}$, and the output power, $P_{out}$, are found from equations on the total power at the bridge $P_B$, (3), and the total power at input of the low-noise amplifier, $P_{LNA}$, (4)

$$dP_B = P_{in} + P_{bias}dS_{31} = d(P_{bias}S_{31}) \quad (3),$$

$$dP_{LNA} = P_{out} + S_{21}dP_{bias} = d(P_{bias}S_{21}) \quad (4).$$

Here right sides are based on data available from the experiment. Assuming that ETF cannot be part of signal power, we set $dS_{31}= 0$ in (3). In turn, the bias power variation cannot be a part of the output signal meaning $dP_{bias}= 0$ in (4). The ratio $P_{out}/P_{in} = Gain$ thus yields

$$Gain = \frac{P_{bias}dS_{21}}{S_{31}dP_{bias}} \quad (5)$$

that is presented in Fig. 3 for a number of bath temperatures. $Gain(T_e)$ can exceed unity within the range of optimum temperatures $T_e = 325\ldots330$ mK growing for lower $T_{bath}$ that corresponds to larger power $P_{bias}$ according to Eqs. (1) and (5). $Gain(T_e)$ reaches its maximum when resistance of the bridge $R_B=Re(Z_B(T_e))$ matches its embedding impedance, $R_s \approx 2.7$ Ω. The EM-model predicts 30% loss of transmission between the bridge and LNA. That is why the gain curves are merging for all $T_{bath}$ at value 0.7 as it should be for a conventional shunted TES in the regime of deep ETF at $R_B \gg R_s$. The discrepancy between measured and theoretical $Z(T_e)$ shown in Fig. 3 can be explained by inaccuracy of determining $P_{in}$ and some unknown dissipation in the resonator and the silicon substrate even at the lowest ("nearly zero") probing power.

The impact of the buffering amplifier, $NEP_{LNA}$, can be written in the following way[7] and presented in Fig. 4:

$$NEP_{LNA} = \sqrt{\frac{k_B T_{LNA}}{2P_{bias}}} \times \frac{1}{S_{1/w}} = \sqrt{\frac{k_B T_{LNA} P_{bias}}{2}} \times \frac{1}{Gain} \quad (6).$$

This equation predicts lower *NEP* for higher $P_{bias}$ due to growing *Gain*. Thus, larger $P_{bias}$ at lower $T_{bath}$ are favorable for both higher gain and higher saturation power, which are always proportional to $P_{bias}$. Here it is worth to repeat that the high bias power, which is limited for MKID due to growing number of dark excitations, is anticipated to be all-over beneficial for the RFTES/EGA detector.

In conclusion, we studied the non-linear microwave impedance of a hafnium microbridge, which can be described with the model of electron gas absorption. We demonstrated that the mew method relates to the electron gas thermometry, so it can be used to design an FDM array-bolometer. This RFTES/EGA technology suits for stratospheric studies at $NEP \approx 10^{-17}$ W/√Hz at saturation powers ~ 1 pW using balloon platforms. Reducing volume of the bridge, the *NEP* below $\approx 10^{-18}$ W/√Hz can be achieved that suites for CMBR-limited operation in space at moderate refrigeration temperatures of 200-300 mK. One may anticipate that the useful *low*-frequency limit of EGA is below its readout frequency.

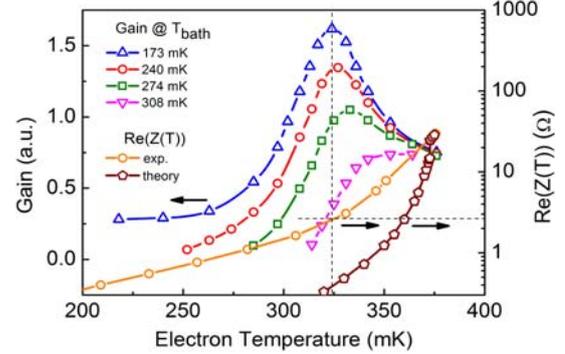

FIG 3. Active impedance $Re(Z(T))$ and gain factor on electron gas temperature for different bath temperatures. Impedance calculated using the Mattis-Bardin theory, $T_{lattice} \approx T_e$, is shown with open pentagons.

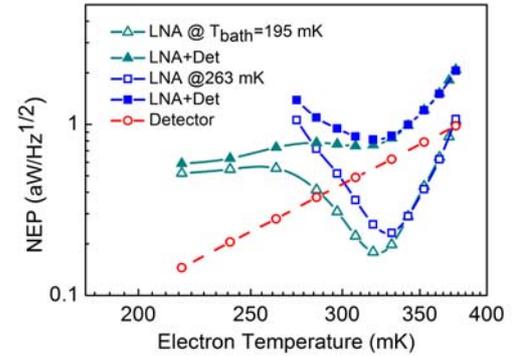

FIG 4. Noise equivalent power for a bridge sized 0.25 μm by 0.25 μm by 25 nm (circles). Effect of buffering LNA (solid boxes and solid triangles) corresponds to amplifier noise $T_{LNA} \approx 1$ K and calculated for two different bath temperatures from Fig. 2.

The *high*-frequency limit is set by the conductive antenna, presumably in the near-infrared. Perhaps, a direct illumination of the airy-size bridge, similar to a pad absorber, can push this limit up. Finally, argue that the RFTES/EGA technology, in general, combines advantages of low *NEP*, higher saturation power, and ease-of-use, which makes it competitive among the state-of-the-art devices from its class.

This work was supported, in part of experimental study, by grant 17-19-01786 from Russian Science Foundation; and, in part of numerical simulations, by Increase Competitiveness Program of NUST «MISIS» K2-2018-015 from the Ministry of Education and Science of the Russian Federation.

[1] S. F. Lee, J. M. Gildemeister, W. Holmes, A. T. Lee, and P. L. Richards, *Applied Optics*, vol. 37, pp. 3391-3397 (1998).
[2] K. D. Irwin and G. C. Hilton," *Topics Appl. Phys.*, vol. 99, pp. 63–149 (2005).
[3] W. Holland, *Proc. SPIE* , vol. 6275, 62751E (2006).
[4] T. M. Lantinga, H. Cho, J. Clarke, M. Dobbs, A. T. Lee, M. Lueker, P. L. Richards, A. D. Smith, H. G. Spieler, *Millimeter Submillimeter Detectors Astronomy*, vol. 4855, pp. 172–181 (2003).

[5] D. K. Day, H. G. LeDuc, B. A. Mazin, A. Vayonakis, and J. Zmuidzinas, *Nature*, vol. 425, pp. 817–821 (2003).
[6] D. K. Day, H. G. LeDuc, B. A. Mazin, A. Vayonakis, and J. Zmuidzinas, *Nature*, vol. 425, pp. 817–821 (2003).
[7] J. Gao, Ph.D thesis, California Institute of Technology (2008).
[8] J. Zmuidzinas, *Annu. Rev. Condens. Matter Phys.*, vol. 3, p. 169 (2012).
[9] A. V. Sergeev and M. YU. Reiser, *Intl. J. Mod. Phys. B*, vol. 10, p. 635 (1996).
[10] F. C. Wellstood, C. Urbina, J. Clarke, *Phys. Rev. B*, vol. 49, pp. 5942-5955 (1994).
[11] B. S. Karasik, W. R. McGrath, H. G. LeDuc and M. E. Gershenson, *Supercond. Sci. Technol.*, vol. 12, pp. 745-747 (1999).
[12] M. E. Gershenson, D. Gong, and T. Sato, *Appl. Phys. Lett.*, vol. 79, Art. no. 2049 (2001).
[13] B. S. Karasik, D. Olaya, J. Wei et al., *IEEE Trans. Appl. Supercond.*, vol. 17, pp. 293-297 (2007).
[14] B. S. Karasik and R. Cantor, *Appl. Phys. Lett.*, vol. 98, 193503 (2011).
[15] D. Golubev and L.S. Kuzmin, *Journal of Applied physics*, vol. 89, p. 6464 (2001).
[16] M. A. Tarasov, L. S. Kuzmin, V. S. Edelman, S. Mahashabde and P. de Bernardis, *IEEE Trans. Appl. Supercond.*, vol. 21, no. 6, p. 3635 (2011).
[17] S. V. Shitov, A. A. Kuzmin, M. Merker, V. I. Chichkov, A. V. Merenkov, A. B. Ermakov, A. V. Ustinov, M. Siegel, *IEEE Trans. Appl. Supercond.*, 27, 4, 2100805 (2017).
[18] A. A. Kuzmin, S. V. Shitov, A. Scheuring, J. M. Meckbach, K. S. Il'in, S. Wuensch, A. V. Ustinov, M. Siegel, *IEEE Trans. THz Sci Technol.*, vol. 3, pp. 25-31 (2013).
[19] S. V. Shitov, N. N. Abramov, A. A. Kuzmin, M. Merker, M. Arndt, S. H. Wuensch, K. S. Ilin, E. V. Erhan, A. V. Ustinov, M. Siegel, *IEEE Trans. Appl. Supercond.*, 25, 3, 2101704 (2015).
[20] A. A. Kuzmin, A. D. Semenov, S. V. Shitov, M. Merker, S. H. Wuensch, A. V. Ustinov, M. Siegel, *App. Phys. Lett.*, vol. 111, pp. 042601 (2017).
[21] M. D. Audley, A. Detrain, L. Ferrari, et. al., *Proc. 23rd Int. Symp. Space Terahertz Technology, ISSTT 2012*, pp. 189-197 (2012).
[22] A. V. Merenkov, V. I. Chichkov, A. B. Ermakov, A. V. Ustinov and S. V. Shitov, *IEEE Trans. Appl. Supercond.*, vol. 28, no.7 (2018).
[23] D. C. Mattis, J. Bardeen, *Phys. Rev.* vol. 111, pp. 412–417, 1958.
[24] A. D. Semenov, G. N. Gol'tsman, R. Sobolewski, *Supercond. Sci. Technol.*, 15, R1-R16 (2002).
[25] Jhy-Jiun Chang and D. J. Scalapino, *Phys. Rev. B*, 15, 2651 (1977).
[26] https://nanoscience.oxinst.com/products/cryofree-dilution-refrigerators/triton
[27] http://www.awrcorp.com/products/ni-awr-design-environment/microwave-office
[28] A. F. Andreev, *Sov. Phys. JEPT*, vol. 19, p. 1228 (1964).